\documentclass[12pt,preprint]{aastex}
\begin{document}

\newcommand{\lsr}{LSR~1610\,--\,0040}
\newcommand{\twom}{2MASS\,0532+8246}
\newcommand{\twomfl}{2MASS\,1439+1929}
\newcommand{\denis}{DENIS~0205\,--\,1159\,AB}


\title{The First High-Resolution Spectra of 1.3 L Subdwarfs}


\author{A. Reiners\altaffilmark{1,^\star} \and G. Basri}
\affil{Astronomy Department, University of California, Berkeley, CA 94720
\email{[areiners, basri]@astron.berkeley.edu}}
\altaffiltext{1}{Hamburger Sternwarte, Universit\"at Hamburg, Gojenbergsweg 112, D-21029 Hamburg, Germany}
\altaffiltext{$^\star$}{Marie Curie Outgoing International Fellow}

%\altaffiltext{1}{Visiting Astronomer, Cerro Tololo Inter-American Observatory.
%CTIO is operated by AURA, Inc.\ under contract to the National Science%Foundation.}
%\altaffiltext{2}{Society of Fellows, Harvard University.}
%\altaffiltext{3}{present address: Center for Astrophysics,
%    60 Garden Street, Cambridge, MA 02138}
%\altaffiltext{4}{Visiting Programmer, Space Telescope Science Institute}
%\altaffiltext{5}{Patron, Alonso's Bar and Grill}

\begin{abstract}
  
  We present the first high-resolution ($R \approx 31\,000$) spectra
  of the cool sdL \twom, and what was originally identified as an
  early-type L subdwarf (sdL) \lsr. Our work, in combination with
  contemporaneous work by Cushing and Vacca, makes it clear that the
  latter object is more probably a mid-M dwarf with an unusual
  composition that gives it some sub-dwarf spectral features.  We use
  the data to derive precise radial velocities for both objects and to
  estimate space motion; both are consistent with halo kinematics. We
  measure the projected rotational velocities, revealing very slow
  rotation for the old sd?M6 object \lsr.  \twom\ exhibits rapid
  rotation of $v\,\sin{i} = 65 \pm 15$\,km\,s$^{-1}$, consistent with
  the behavior of L dwarfs. This means that the braking time for L
  dwarfs is extremely long, or that perhaps they never slow down.
  
  A detailed comparison of the atomic Rb and Cs lines to spectra of
  field L dwarfs shows the spectral type \twom\ is consistent with
  being mid- to late-L. The Rb~I and K~I lines of \lsr\ are like an
  early-L dwarf, but the Cs~I line is like a mid-M dwarf.  The
  appearance of the Ca~II triplet in absorption in this object is very
  hard to understand if it is not as least as warm as M6. We explain
  these effects in a consistent way using a mildly metal-poor mid-M
  model. M subdwarfs have weak metal-oxides and enhanced
  metal-hydrides relative to normal M dwarfs. \lsr\ exhibits
  metal-hydrides like an M dwarf but metal-oxides like a subdwarf. The
  same explanation that resolves the atomic line discrepancy explains
  this as well.
  
  Our spectra cover the spectral region around a previously
  unidentified absorption feature at 9600\,\AA, and the region around
  9400\,\AA\ where detection of TiH has been claimed. We identify the
  absorption around 9600\,\AA\ as due to atomic lines of Ti and a
  small contribution of FeH, but we cannot confirm a detection of TiH
  in the spectra of cool L subdwarfs. In \twom, both metal-oxides
  \emph{and} metal-hydrides are extremely strong relative to normal L
  dwarfs.  It may be possible to explain the strong oxide features in
  \twom\ by invoking effects due to inhibited dust formation. High
  resolution spectroscopy has aided in beginning to understand the
  complex molecular chemistry and spectral formation in
  metal-deficient and ultracool atmospheres, and the properties of
  early ultralow-mass objects.

\end{abstract}
%% subject header.  Objects should be in the appropriate "individual"
%% headers (e.g. quasars: individual, stars: individual, etc.) with the
%% additional provision that the total number of headers, including each
%% individual object, not exceed six.  The \objectname{} macro, and its
%% alias \object{}, is used to mark each object.  The macro takes the object
%% name as its primary argument.  This name will appear in the paper
%% and serve as the link's anchor in the electronic edition if the name
%% is recognized by the data centers.  The macro also takes an optional
%% argument in parentheses in cases where the data center identification
%% differs from what is to be printed in the paper.
%\keywords{globular clusters: general ---
%globular clusters: individual(\objectname{NGC 6397},
%\object{NGC 6624}, \objectname[M 15]{NGC 7078},
%\object[Cl 1938-341]{Terzan 8})}

\newpage

\section{Introduction}

The most immediate relics of the early Galaxy are the cool subdwarfs
of spectral type sdK and later -- lifetimes of such cool stars are
well in excess of the age of the Galaxy. Their metal-deficient
atmospheres make them appear hotter than solar-metallicity
main-sequence stars of the same mass, which in turn renders them
``subluminous'' \citep{Kuiper39}.

The spectral classification of solar abundance K- and M-dwarfs is well
understood, and is entirely due to changes in effective temperature.
Classification of M-subdwarfs is based on the metallicity sensitive
ratio of absorption bands of metal-oxides and metal-hydrides
\citep{Gizis97, Lepine03a}; metal-hydrides are generally stronger in
metal-deficient atmospheres, while metal-oxides are weaker
\citep[e.g.,][]{Mould76}. In the ultracool atmospheres of late M- and
L-dwarfs, however, refractory grains become an important ingredient
since they are competing with molecules for available metals.
Furthermore, their opacity influences the optical depth, and hence
which part of the atmosphere is visible. Since the formation and
distribution of dust grains is not well understood even in solar
abundance stars, this is an even more severe problem in the
classification of metal-deficient L-subdwarfs.

The last two years have seen the first discoveries of L-type
subdwarfs.  These ultracool dwarfs exhibit colors too red for an
M-type object, and do not fit on the expanded subdwarf classification
scheme of \cite{Lepine03a}. Another argument for surface temperatures
lower than M-dwarf temperatures is the extremely strong absorption
lines of alkali atoms as K~I, Na~I, and Rb~I. The first L-type
subdwarf, 2MASS~J05325346+8246465 (hereafter \twom), was discovered by
\citet[][hereafter B03]{Burgasser03}. Its spectrum is very similar to
the L7 dwarf \denis, exhibiting strong alkali lines \citep[we use the
optical classification scheme of][]{Kirkpatrick99}.  \twom\ shows
enhanced metal hydride bands, but in contrast to M-subdwarfs it also
has strongly enhanced bands of TiO. The first proposed early-type
L-subdwarf, \lsr, was discovered by \citet[][hereafter
L03]{Lepine03b}.  It also exhibits strong Rb~I absorption and L03
report enhanced CaH and TiO bands.  Metal hydrides FeH and CrH,
however, are relatively weak. We conclude in this paper that \lsr\ is
not really an L subdwarf, but in the M temperature range. The same
conclusion is reached in a paper based on lower resolution spectra
that was posted during our refereeing process by \citet[][hereafter
CV05]{CV05} . The spectral peculiarities of these unusual objects
provide a preview of the complex chemical processes occurring in cool
atmospheres of metal-deficient L-dwarfs. The list of subdwarfs of
spectral type L and late M is constantly growing; \citet{Burgasser04a}
discovered the L-subdwarf LSR~$1626 + 3925$, and two late M-type
subdwarfs at the end of the subdwarf classification scheme with
features very similar to the L-type subdwarfs have been reported by
\citet[][SSSPM~J1013\,--\,1356]{Scholz04a} and
\citet[][SSSPM~J1444\,--\,2019]{Scholz04b}.

The presumably large age of M- and L-subdwarfs also makes them ideal
tracers of the rotational evolution of late-type objects. Cool stars
of spectral type K and M are known to suffer rotational braking on a
relatively short timescale (1 Gyr), but no projected rotation velocity
below $v\,\sin{i} = 10$\,km\,s$^{-1}$ has been reported in an L-dwarf
\citep{Mohanty03, Bailer04}. Whether this is due to extremely long
braking times or the lack of rotational braking at all in L-dwarfs may
be answered by observations of the rotation of the oldest objects of
very late spectral type.

So far, no high-resolution spectrum of an L-type subdwarf has been
available to investigate the details of line strengths, take a close
look for unidentified features detected in L-subdwarf spectra,
\citep[among them the tentative detection of TiH,][]{Burgasser04b},
and to investigate rotation velocities. In this paper, we present the
first high resolution spectra of the late-type sdL \twom\ and the peculiar 
sd?M \lsr.  We describe our observations in \S~2.  Radial
velocities are calculated in \S~3, where we also derive new
constraints on the space motion on the basis of the accurate radial
velocities. Surface rotation is investigated in \S~4, and individual
spectral features are discussed in \S~5.

\section{Observations}

Both targets were observed with HIRES at Keck~I on March 2, 2005,
after the detector upgrade. The spectral sensitivity and coverage are
substantially improved after the upgrade, allowing us to obtain
high-resolution spectra of these faint objects, especially in the CCD
infrared. With the new CCD mosaic we were able to cover the spectral
region from 5700\,\AA\ to 1\,$\mu$m in 25 spectral orders. In the red
part of the spectrum, however, spectral coverage is still incomplete.
The CCD mosaic is made of three chips; two spectral orders fall onto
the gaps and could not be observed.

We obtained a signal-to-noise ratio (SNR) better than 30 around
8000\,\AA\ in a 40\,min exposure of \lsr. The much fainter \twom\ was
observed for 45\,min yielding much lower a SNR in the optical
wavelength region.  However, due to the improved sensitivity in the
red CCD, we achieved a SNR well above ten in the near infrared region
redward of 8500\,\AA.

We positioned the CCD mosaic in order to obtain the spectral features
most important in late-type stars and brown dwarfs. Our spectra
include the important alkali lines of Rb~I and Cs~I, the TiO band at
8450\,\AA\, the prominent FeH and CrH bands at $1\mu$m, the TiH region
at 9400\,\AA, and the unidentified spectral feature around 9600\,\AA\ 
(B03). On the other hand, we were not able to cover the Na~I resonance
doublet in the same exposure, and we miss the FeH and CrH bands at
8600\,\AA\ and 8700\,\AA\ since they fall in the gap between two of
the CCDs.  The TiO band at 7040\,\AA\ and the K resonance lines are at
the very edge of our spectra.

\section{Radial velocities and space motion}

\begin{deluxetable}{lc}
  \tablecaption{\label{tab:LSR1610} Parameters of \lsr}
  \tablewidth{0pt}
  \tablehead{Parameter & Value}
  \startdata
  R.A. (J2000.0)  & $16^h10^m28\fs85$\\
  Decl. (J2000.0)  & $-00\degr40\arcmin53\farcs0$\\
  $\mu$\tablenotemark{a}  & $1\farcs46$ yr$^{-1}$\\
  $\theta$\tablenotemark{a} & $212\fdg0$\\
  Distance\tablenotemark{a}  & $16 \pm 4$\,pc\\
  2 MASS $J$ &   $12.91 \pm 0.02$\,mag\\
  2 MASS $H$ &   $12.32 \pm 0.02$\,mag\\
  2 MASS $K_s$ & $12.02 \pm 0.03$\,mag\\
  $v_{\rm{rad}}$  & $-95 \pm 1$\,km\,s$^{-1}$\\
  $U$  & $-$\phn$44 \pm \phn8$\,km\,s$^{-1}$\\
  $V$  & $-111 \pm 27$\,km\,s$^{-1}$\\
  $W$  & $-\phn51 \pm \phn2$\,km\,s$^{-1}$\\
  $v\,\sin{i}$ & $<5$\,km\,s$^{-1}$\\
  \enddata
  \tablenotetext{a}{\cite{Lepine03b}}
\end{deluxetable}

\begin{deluxetable}{lc}
  \tablecaption{\label{tab:2MASS0532} Parameters of \twom}
  \tablewidth{0pt}
  \tablehead{Parameter & Value}
  \startdata
  R.A.\tablenotemark{a} & $5^h32^m53\fs46$\\
  Decl.\tablenotemark{a} & $+82\degr46\arcmin46\farcs5$\\
  $\mu$\tablenotemark{b}    & $2\farcs60 \pm 0\farcs015$\,yr$^{-1}$\\
  $\theta$\tablenotemark{b} & $130\fdg0 \pm 1\fdg8$\\
  Distance\tablenotemark{c} & $20 \pm 10$\,pc\\
  2 MASS $J$     &   $15.81 \pm 0.06$\,mag\\
  2 MASS $H$     &   $14.90 \pm 0.10$\,mag\\
  2 MASS $K_s$   &   $14.92 \pm 0.15$\,mag\\
  $v_{\rm{rad}}$ &   $-172 \pm \phn1$\,km\,s$^{-1}$\\
%  $U$            & $-$\phn26\,km\,s$^{-1}$\\
%  $V$            & $-285$\,km\,s$^{-1}$\\
%  $W$            & \phn\phs$38$\,km\,s$^{-1}$\\
  $v\,\sin{i}$   &  \phn\phs$65 \pm 15$\,km\,s$^{-1}$\\
  \enddata
  \tablenotetext{a}{Epoch 1999 March 1 (UT), \cite{Burgasser03}}
  \tablenotetext{b}{\cite{Burgasser03}}
  \tablenotetext{c}{estimate from optical spectrum, \cite{Burgasser03}}
\end{deluxetable}

We calculate the radial velocities of \lsr\ and \twom\ by comparison
to the radial velocity of Gl\,406 \cite[$v_{\rm rad} =
19\pm1$km\,s$^{-1}$,][]{Martin97}, which was observed during the same
night. Radial velocities are measured from a cross correlation with
the spectrum of Gl\,406 and applying the barycentric corrections.  We
give all parameters of \lsr\ and \twom\ in Tables\,\ref{tab:LSR1610}
and \ref{tab:2MASS0532}, respectively.

\subsection{\twom}
\label{sect:vradtwom}

To calculate the radial velocity of \twom, we employ two spectral
orders covering the regions around 9200\,\AA\ and before 1$\mu$m where
strong features of FeH are dominant. After barycentric correction, we
calculate a relative radial velocity between \twom\ and Gl\,406 of
$\Delta v_{\rm rad} = -191 \pm 1$\,km\,s$^{-1}$, i.e., $v_{\rm rad} =
-172 \pm 1$\,km\,s$^{-1}$. This result is within 2$\sigma$ of the
value reported by B03 ($v_{\rm rad} = -195 \pm 11$\,km\,s$^{-1}$). In
fact, B03 observe a stronger blueshift when they derive $v_{\rm rad}$
from optical Cs~I and Rb~I lines. The radial velocity they derive from
infrared K~I lines at 1.1690/1.1773\,$\mu$m, $<$$v_{\rm{rad}}$$>\,=
-175 \pm 17$\,km\,s$^{-1}$, matches much better with our value from high
resolution spectra. 

In order to estimate the space motion of \twom, we adopt the
parameters given by B03.  Assuming that the spectroscopically derived
I$_C$-magnitude of \twom\ is similar to late-type L dwarfs, they
estimate a distance of $d =$~10\,--\,30\,pc. The radial velocity we
derive from our data only yields small corrections to the space motion
vector they give, the new result is $[U, V, W] = [-26, -285,
38]$\,km\,s$^{-1}$.  However, due to the uncertain distance, the space
motion is poorly constrained and we do not include it in
Table\,\ref{tab:2MASS0532}.  Nevertheless, the likelihood of a very high
$V$-velocity \twom\ does significantly exceed the 2$\sigma$ limit
for disk stars, as was mentioned by B03.

\subsection{\lsr}
\label{sect:vradlsr}

For \lsr, we use three spectral orders around 8000\,\AA. We checked
the reliability of our radial velocity determination by measuring
$v_{\rm rad}$ for GJ\,299, GJ\,1227 and GJ\,1111 from observations
taken during the same night.  In all three cases our results show
excellent agreement with the values reported in \cite{Delfosse98} and
\cite{Mohanty03}. The relative velocity difference between \lsr\ and
Gl\,406 as calculated from the spectra is $\Delta v_{\rm rad} = -113.7
\pm 0.2$\,km\,s$^{-1}$ after barycentric correction. Our uncertainty
is dominated by the star's position in the slit, and we estimate a
systematic uncertainty of 1\,km\,s$^{-1}$. Thus, for \lsr\ we
calculate a radial velocity of $v_{\rm rad} = -95 \pm
1$\,km\,s$^{-1}$. 

This results is not in agreement with the radial velocity of $v_{\rm
  rad} = -130 \pm 15$\,km\,s$^{-1}$ reported by L03.  However, B03
calculated $v_{\rm rad}$ for \twom\ and found a difference between
their values of $v_{\rm rad}$ derived from optical and from infrared
wavelength regions (see Sect.\,\ref{sect:vradtwom}).  Using low
resolution data at optical wavelengths they obtain a radial velocity
that is about 20\,km\,s$^{-1}$ more blueshifted than $v_{\rm rad}$
derived from infrared regions. As was mentioned above, our results
agree with the result they get from the infrared data. We suggest that
such an offset is also the reason for the mismatch between the radial
velocity derived in low-resolution data at optical wavelengths by L03
and the value we derive from our high-resolution data. If we correct
the radial velocity reported by L03 by the +20\,km\,s$^{-1}$ reported
by B03, the corrected value is $v_{\rm rad} = -110 \pm
15$\,km\,s$^{-1}$, which is consistent with the value of $v_{\rm rad}
= -95 \pm 1$\,km\,s$^{-1}$ we find.

To calculate the space motion vector [$U, V, W$], we first tried to
reproduce the vector reported by L03. We use the IDL procedure
{\ttfamily gal\_uvw}\footnote{\ttfamily
  http://idlastro.gsfc.nasa.gov/contents.html} and the coordinates,
proper motion and radial velocity given in L03. However, we do not
derive the same space motion (while we do get the identical vector B03
derive for \twom). Our result is $[-72, -117, -71]$\,km\,s$^{-1}$,
while L03 report $[-111, -85, -24]$\,km\,s$^{-1}$ (they also write
slightly different values in their Table\,1 and their Sect.\,5).
Including the new radial velocity for \lsr\ we derive our final result
of $[U, V, W] = [-44, -111, -51]$\,km\,s$^{-1}$. This places \lsr\ 
inside the $2\sigma$ limit of thick disk stars, as defined by
\cite{Chiba00}. Nevertheless, the $V$-velocity of \lsr\ of $-111 \pm
27$\,km\,s$^{-1}$ still is indicative of halo kinematics, although it
is not statistically significant in comparison to the sample of
\citet{Chiba00}.

\section{Rotation velocities}
\label{sect:vsini}

\begin{figure}
%put the following line into the .eps files if they are redone
%%BoundingBox: 0 0 450 468
  \plottwo{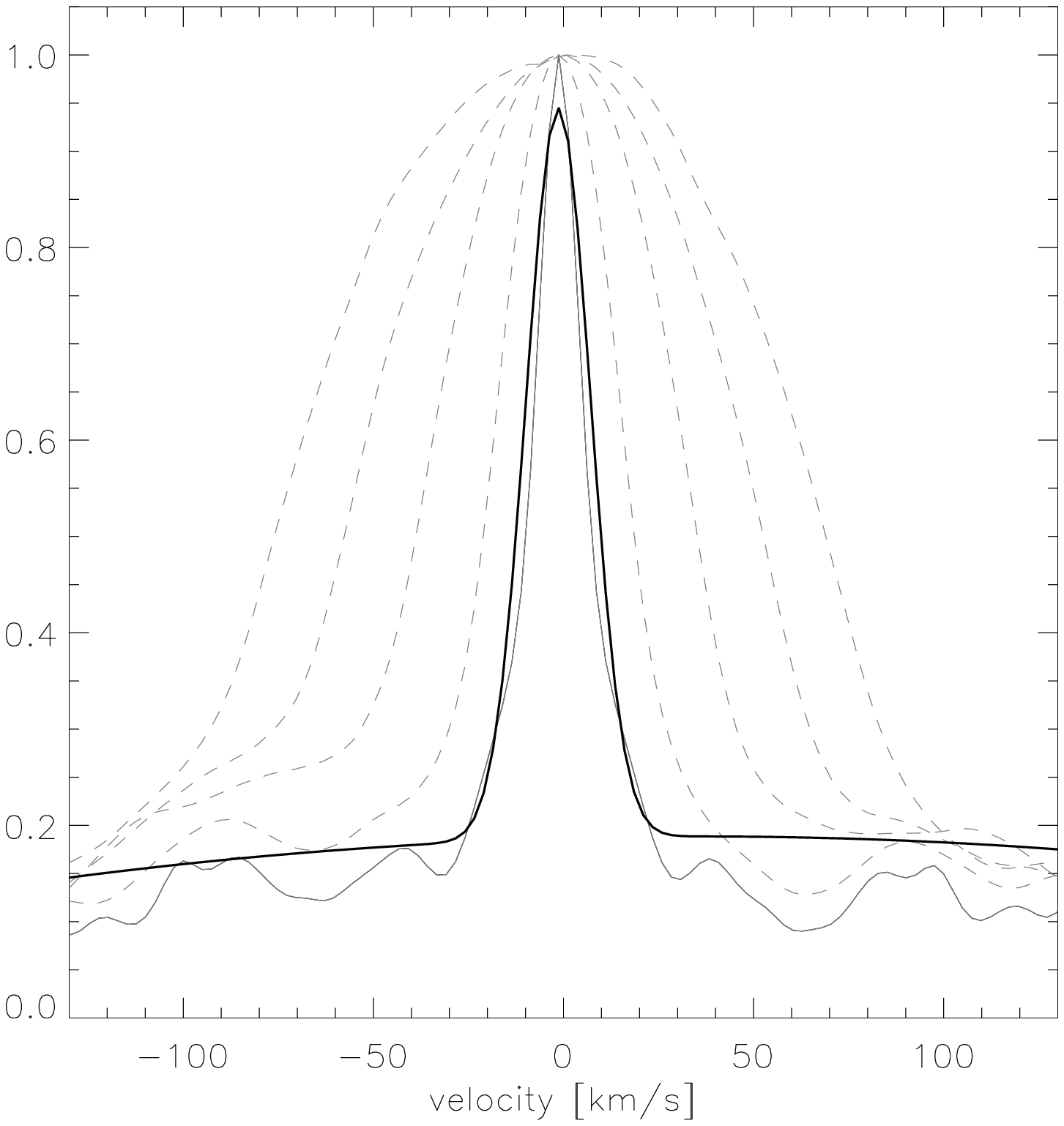}{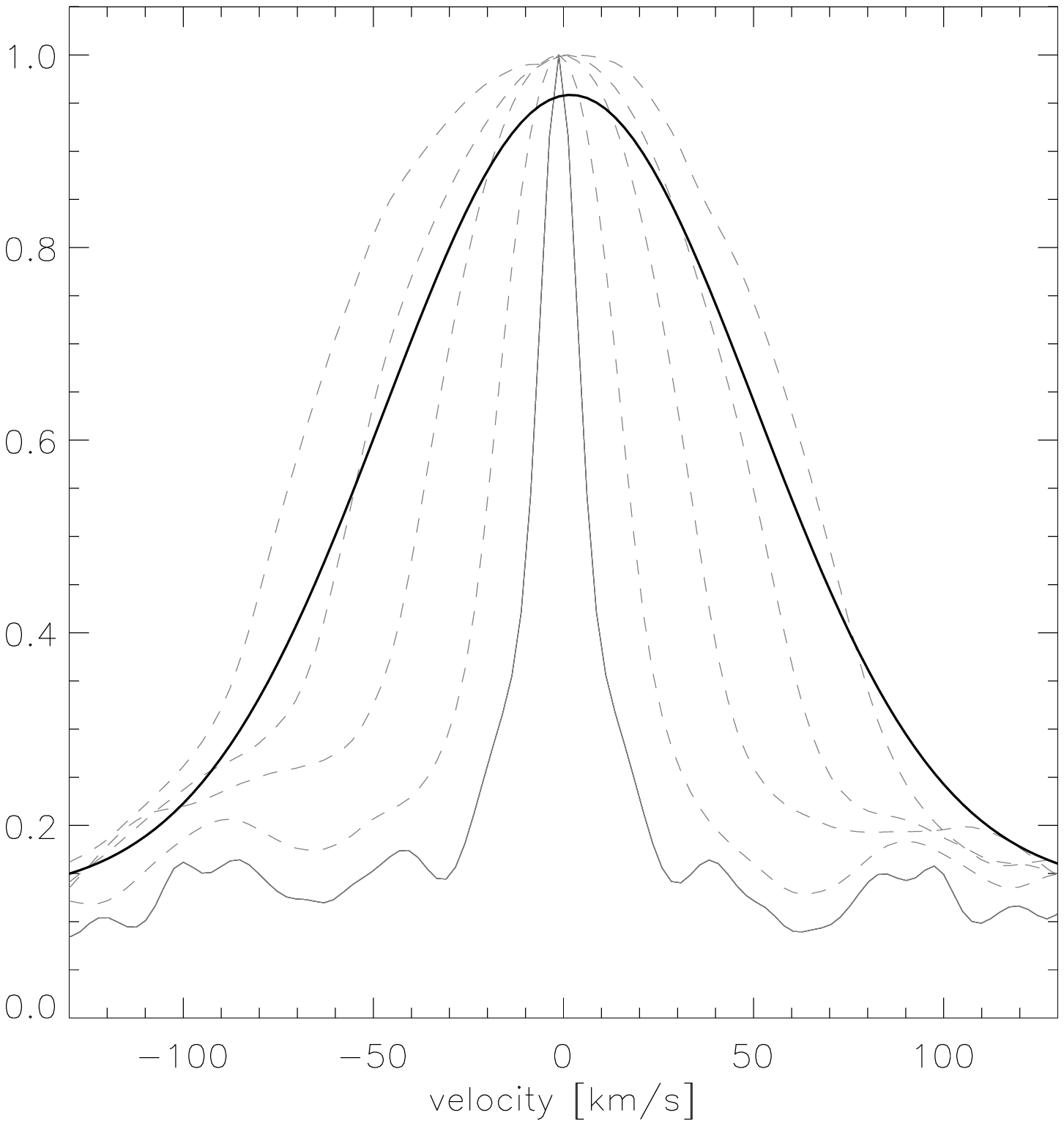}
  \caption{\label{fig:xcorl}Cross correlation function of Gl\.406 and \lsr\ (left panel) and 
    \twom\ (right panel). Correlation functions of the spectrum of
    Gl\,406 with a version of that template spun up by $v\,\sin{i} =
    0, 20, 40 ,60, 80$\,km\,s$^{-1}$, in order of the narrowest to
    widest curves, are overplotted for calibration (dashed lines).}
\end{figure} 

The predominant spectral features visible in L-type spectra are molecular
bandheads and strongly pressure-broadened alkali lines. Neither of
them allow a determination of $v\,\sin{i}$ by directly fitting a
rotationally broadened template to the spectrum, as can be
successfully done in hotter stars where the shape of unblended atomic
lines indicate stellar rotation very precisely. The projected rotation
velocities $v\,\sin{i}$ of ultralow mass objects as late as spectral
type L are usually determined via cross-correlation with a slowly
rotating M- or L-type object used as a template \citep{Basri00}. The
spectral type of the template star should be as close as possible to
the target. One of the latest slow rotating M-dwarfs bright enough to
be employed as a template star is Gl\,406 (M5.5), and that star is
often used for the cross-correlation technique also in L-dwarfs. It is
not entirely clear to what extent spectral differences between the
spectra of M- and L-type dwarfs, e.g.  the severe broadening of alkali
lines in L-dwarfs, could mimic a higher rotation velocity when using
an M-dwarf template to derive $v\,\sin{i}$ in an L-dwarf.
\citet{Mohanty03} discuss this issue in detail. \cite{Bailer04} used
the spectrum of \twomfl\ (L1.0, $v\,\sin{i} \approx 10$\,km\,s$^{-1}$)
to calculate $v\,\sin{i}$ in his L-dwarf sample.  He found no
significant differences between the two values of $v\,\sin{i}$ he
individually derived from a cross-correlation with Gl\,406, and with
\twomfl. This supports the assumption that using an M-dwarf for the
template does in fact yield the correct $v\,\sin{i}$ as well.

The slowest projected rotation velocity measured in an L-dwarf is as
high as $v\,\sin{i} \approx 10$\,km\,s$^{-1}$ \citep{Mohanty03, 
Bailer04}, while mid-M dwarfs are mixed and early-M dwarfs 
usually have rotation velocities below  the detection limit. 
L-dwarfs may not suffer rotational braking at all,
since their neutral atmospheres could inhibit any kind of magnetic 
coupling that is the source of rotational braking in hotter stars 
\citep{Mohanty02}. However, flares have been observed to occur
in L-dwarfs and some are observed to exhibit H$\alpha$ emission 
\citep{Leibert03}. This indicates at least the possibility of 
magnetic braking. The lack of slowly rotating L-dwarfs could
also be an artifact of observing predominantly young objects since
they are simply brighter than older objects of the same mass. In that
case slow rotators might be found among the very oldest L-dwarfs, the
subdwarf population.

We calculated the cross-correlation functions of our target stars 
\lsr\ and \twom\ with Gl\,406 in several spectral orders. Results from
different spectral orders are consistent with each other. The
cross-correlation functions of \lsr\ and \twom\ from the wavelength
region around 9150\,\AA\ are shown in the left and right panels of
Fig.\,\ref{fig:xcorl}, respectively. To calibrate the width of the
cross-correlation functions, we spun up the spectrum of Gl\,406
according to surface velocities $v\,\sin{i} = 20, 40, 60$ and
80\,km\,s$^{-1}$ and calculated the cross-correlation functions with
the unbroadened template of Gl\,406. The results are overplotted in
both panels of Fig.\,\ref{fig:xcorl} as dashed lines.

It is immediately clear from the left panel of Fig.\,\ref{fig:xcorl}
that \lsr\ does not show any signs of a rotation faster than Gl\,406
($v\,\sin{i} \approx 3$\,km\,s$^{-1}$). This can be viewed as support
for the proposition that \lsr\ is more similar to Gl\,406 than an
early L dwarf (which is expected to be a rapid rotator).

One advantage of the cross-correlation technique is its applicability
to spectra of relatively low signal. The SNR of our spectrum of \twom\ 
is much lower because of the faintness of object. Nevertheless, it is
possible to derive reliable cross-correlation functions and determine
the projected rotation velocity as shown in the right panel of
Fig.\,\ref{fig:xcorl}. From that function, we derive a value of
$v\,\sin{i} = 65$\,km\,s$^{-1}$ for \twom. The uncertainty of our
result can be derived from comparing results from different spectral
orders; we estimate an uncertainty of 15\,km\,s$^{-1}$ (rapid rotation 
is more difficult to be precise with). If \twom\  is a relic of the early 
Galaxy, we are led to the conclusion that the braking time is longer than the 
age of the Galaxy or there is no braking at all for such cool objects. 

\section{Spectra of Individual Objects}

\subsection{\lsr}

\begin{figure*}
  \centering \leavevmode 
  \mbox{
    \includegraphics[width=.4\textwidth,bbllx=0,bblly=65,bburx=300,bbury=310]{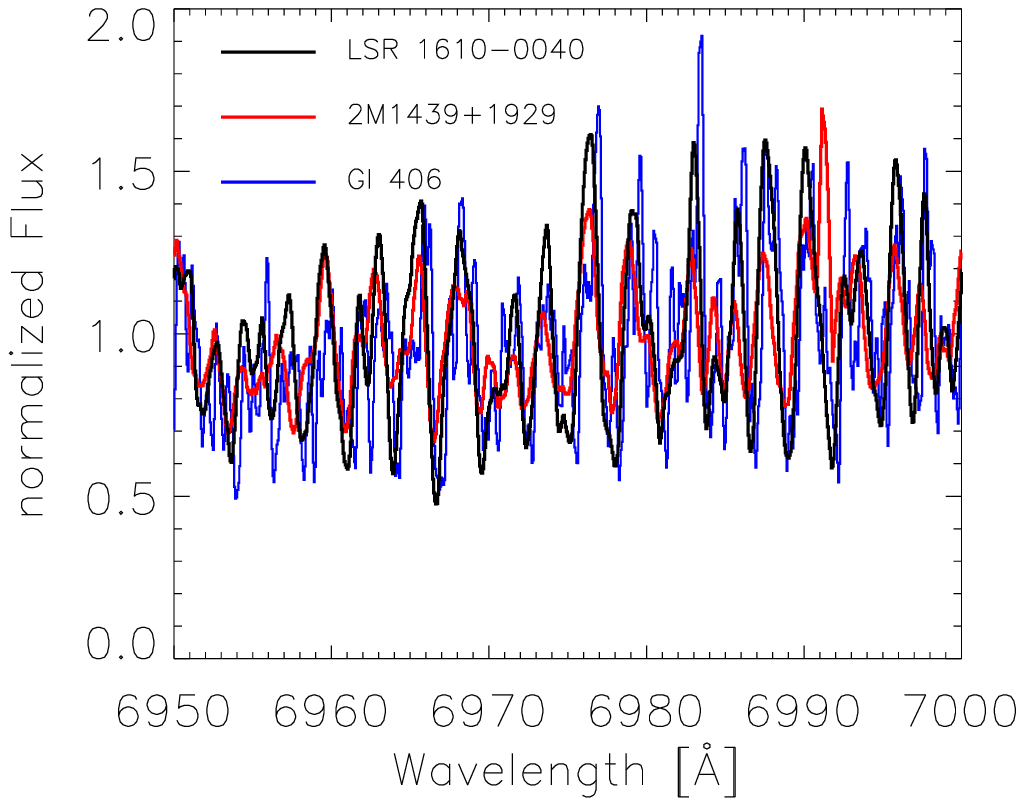}
    \includegraphics[width=.4\textwidth,bbllx=0,bblly=65,bburx=300,bbury=310]{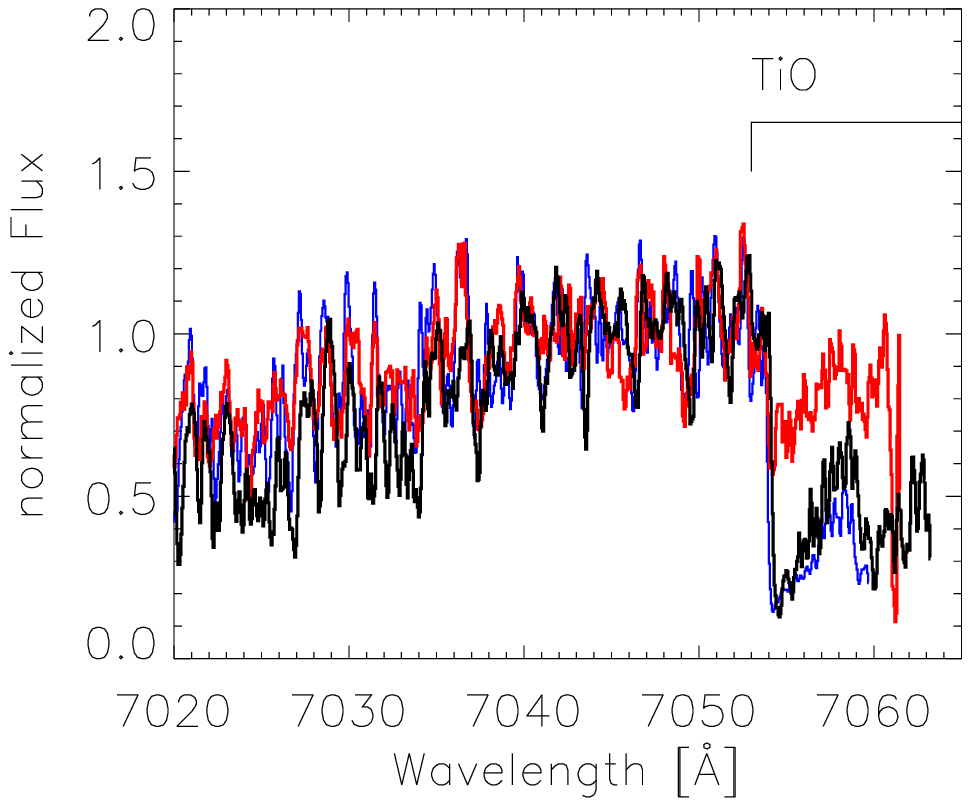}
  }
  \includegraphics[width=.9\textwidth,bbllx=0,bblly=65,bburx=648,bbury=310]{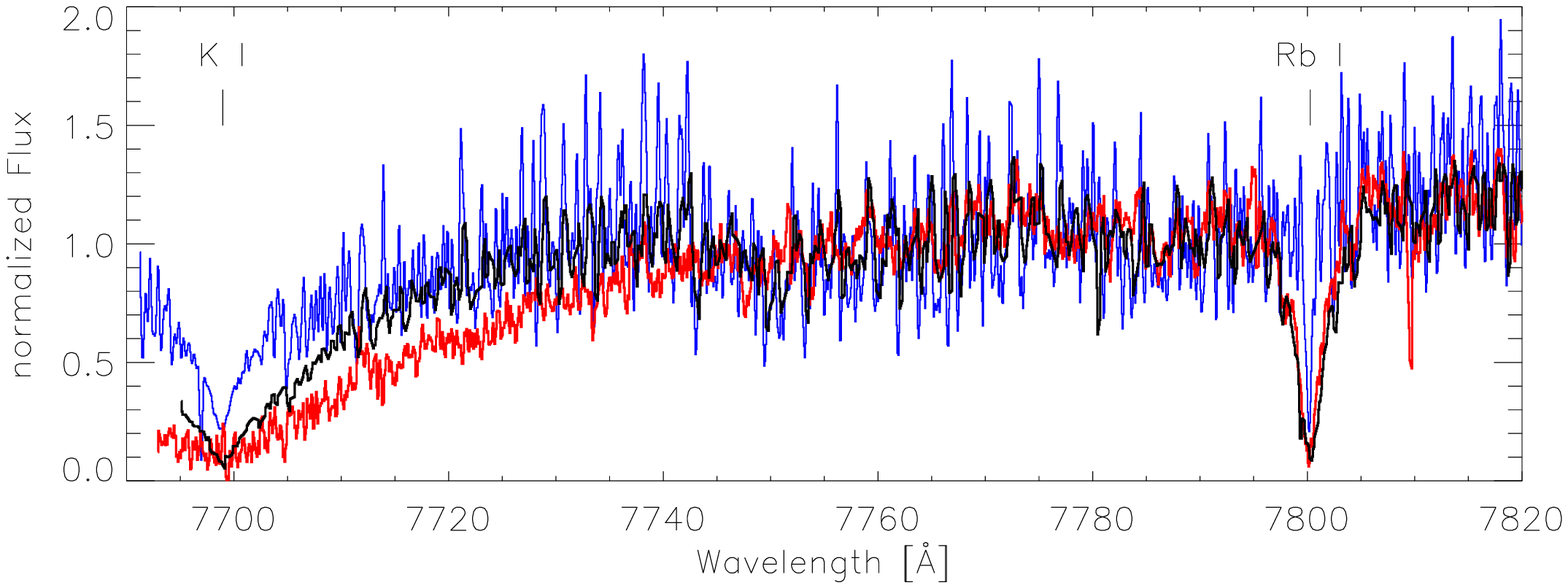}
  \includegraphics[width=.9\textwidth,bbllx=0,bblly=65,bburx=648,bbury=310]{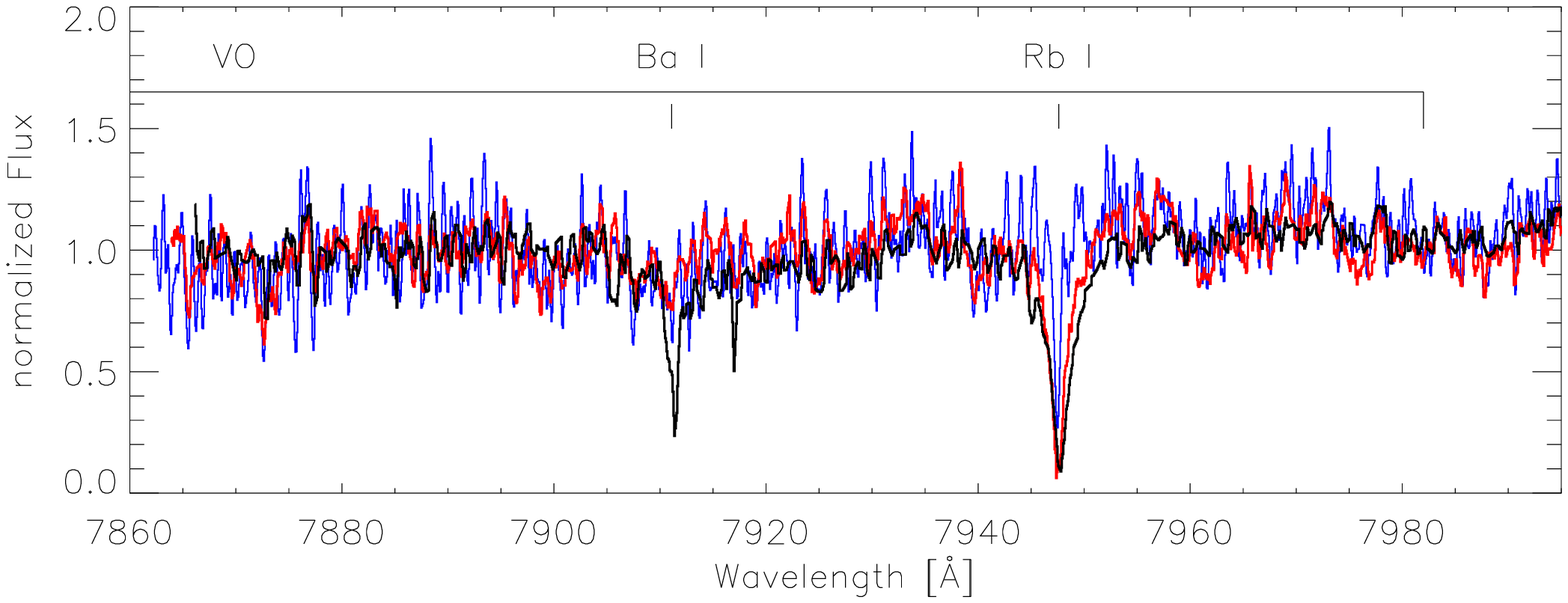}
  \includegraphics[width=.9\textwidth,bbllx=0,bblly=65,bburx=648,bbury=310]{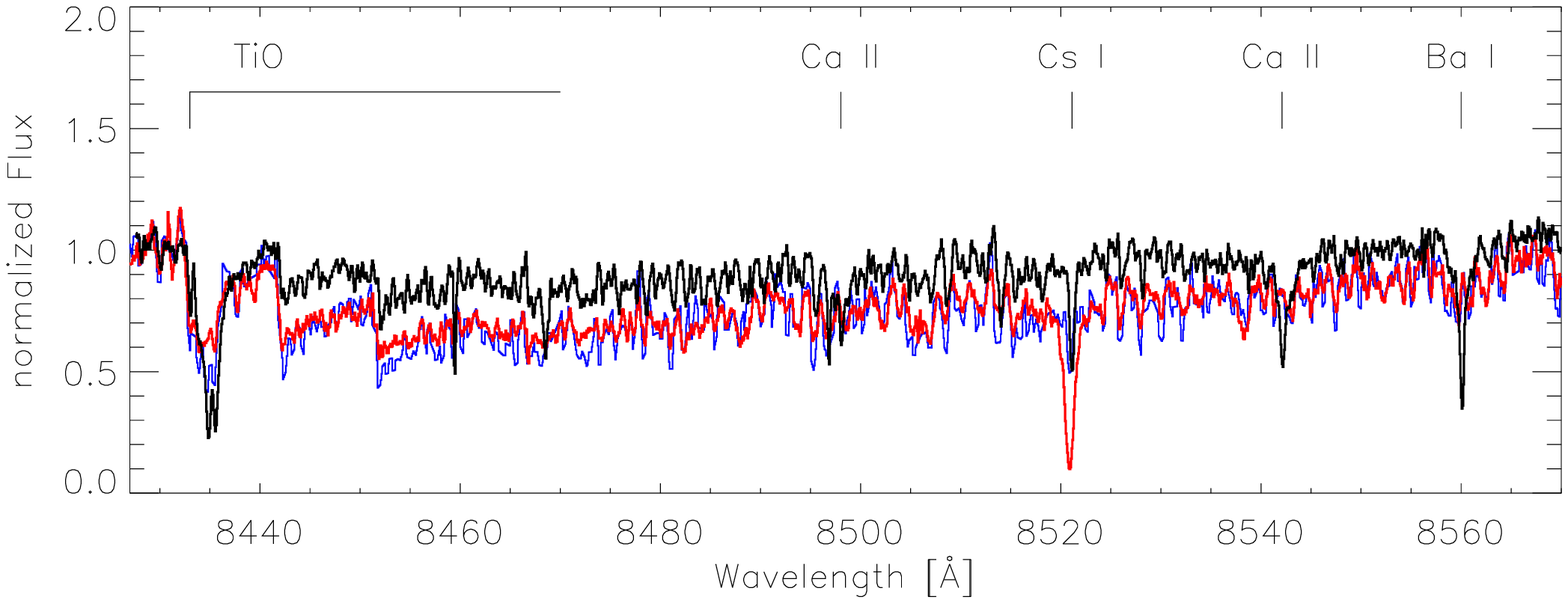}
  \caption{\label{fig:LSR1610_1}Spectra of \lsr\ (black), the L1 field L-dwarf \twomfl\ (red) and the M5.5 dwarf Gl\,406 (blue). The two narrow features in the TiO bandhead at 8435\,\AA\ in the spectrum of \lsr\ are due to atomic Ti. }
\end{figure*} 

\begin{figure*}
  \centering \leavevmode
  \includegraphics[width=.9\textwidth,bbllx=0,bblly=65,bburx=648,bbury=310]{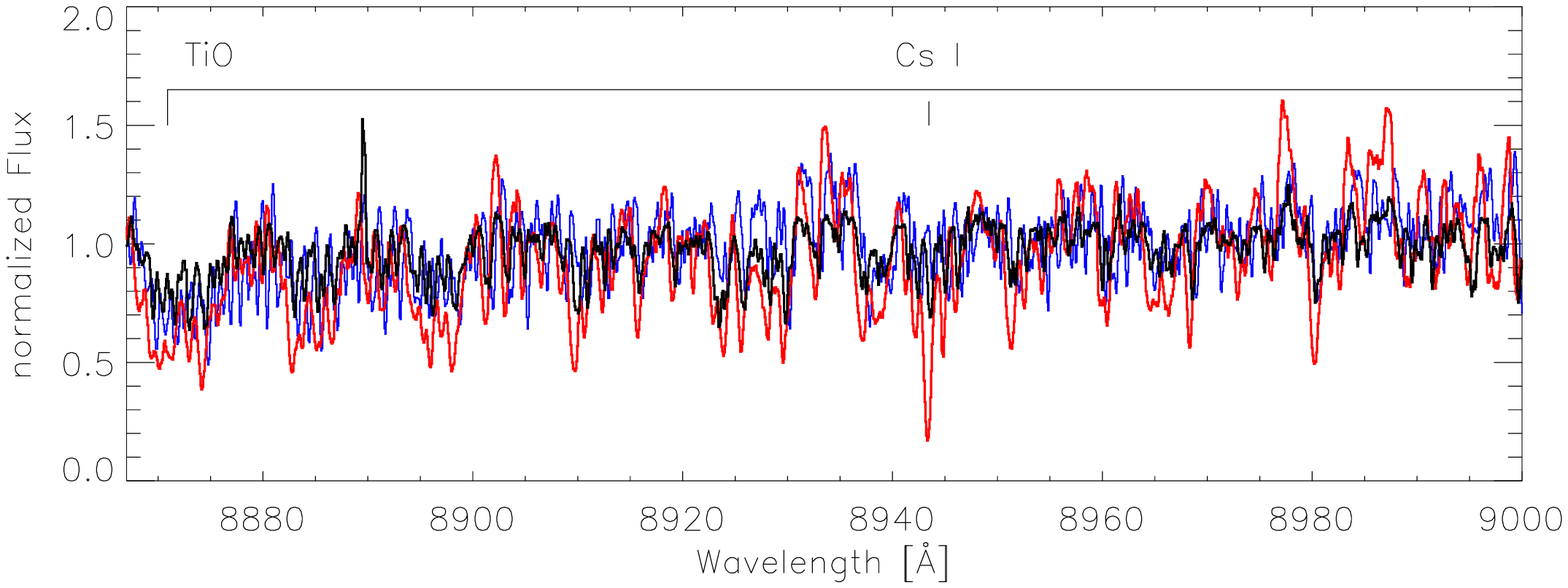}
  \includegraphics[width=.9\textwidth,bbllx=0,bblly=65,bburx=648,bbury=310]{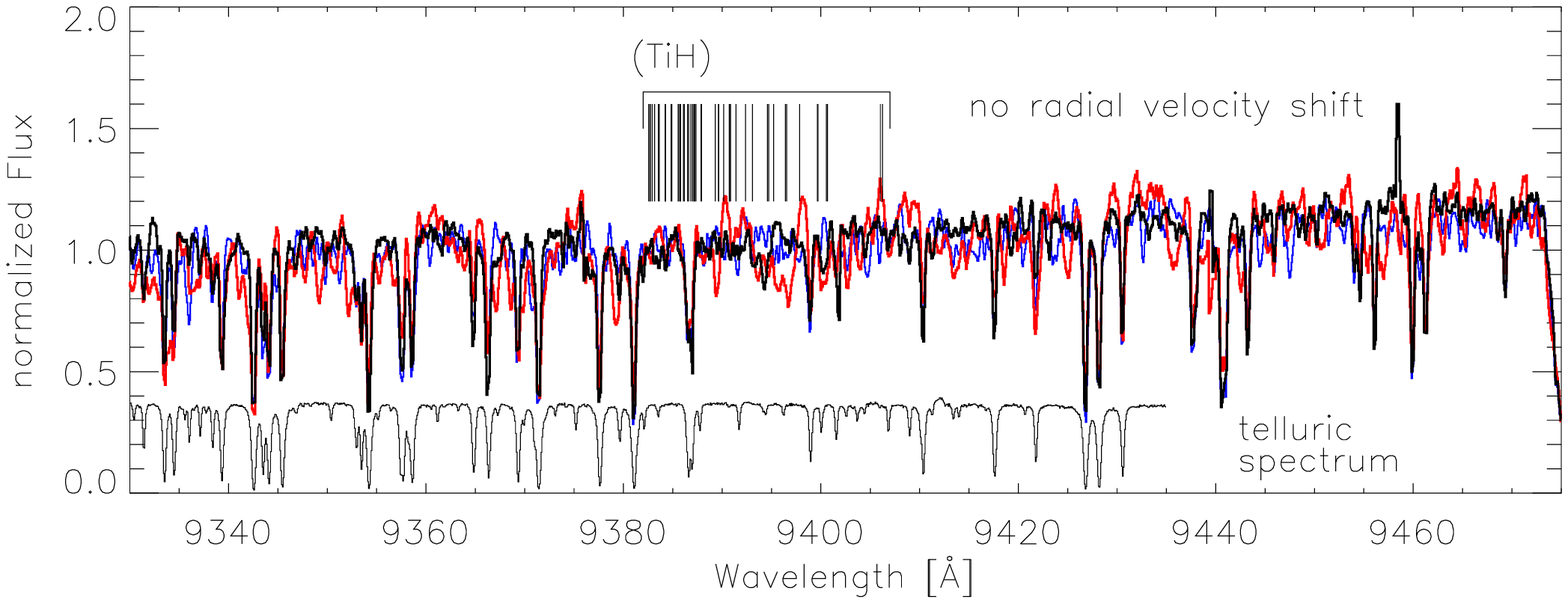}
  \includegraphics[width=.9\textwidth,bbllx=0,bblly=65,bburx=648,bbury=310]{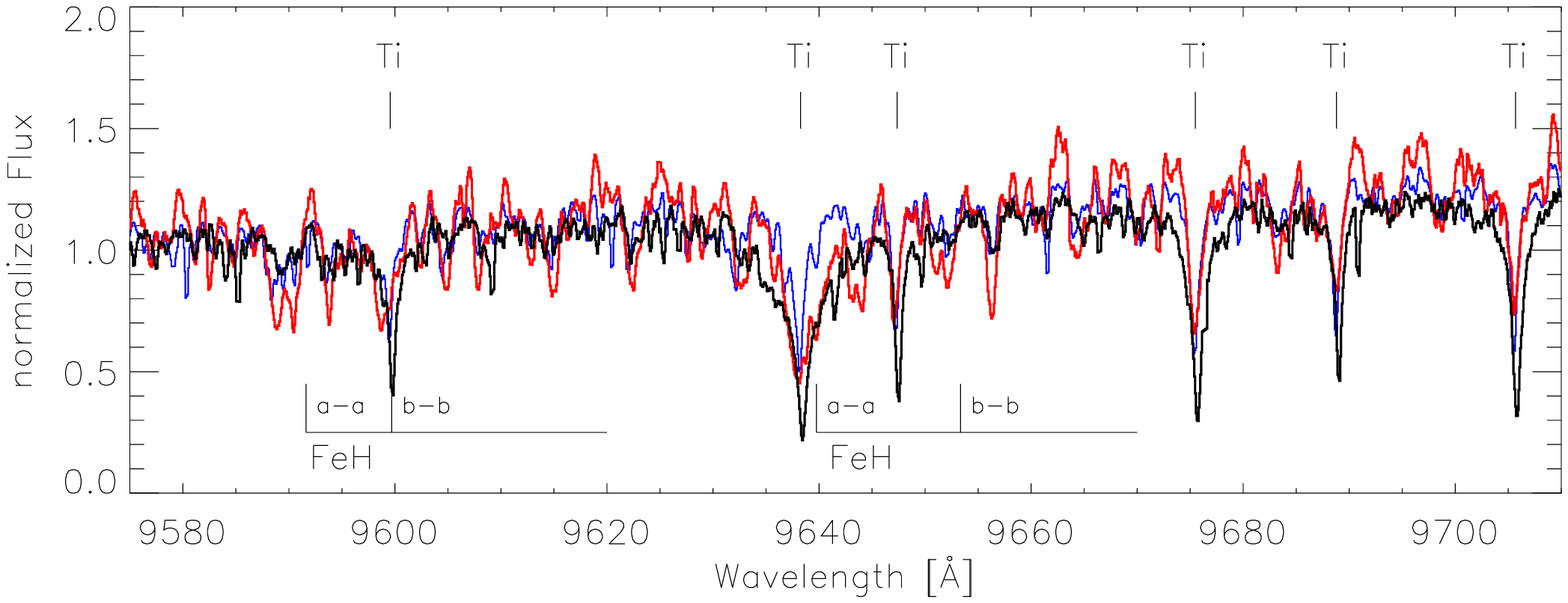}
  \includegraphics[width=.9\textwidth,bbllx=0,bblly=65,bburx=648,bbury=310]{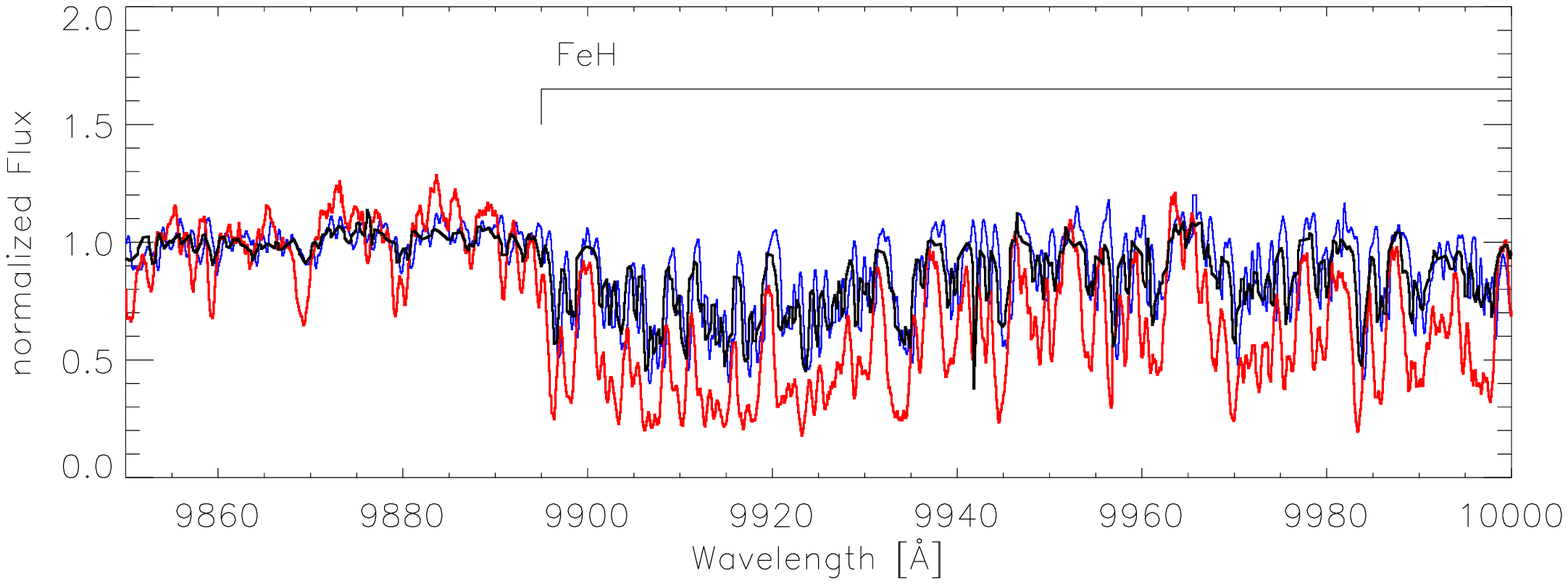}
\caption{\label{fig:LSR1610_2}Spectra of \lsr\ (black), the L1 field L-dwarf \twomfl\ (red) and the M5.5 dwarf Gl\,406 (blue).}
\end{figure*} 

\begin{figure*}
  \centering \leavevmode 
  \mbox{
    \includegraphics[width=.4\textwidth,bbllx=0,bblly=65,bburx=300,bbury=310]{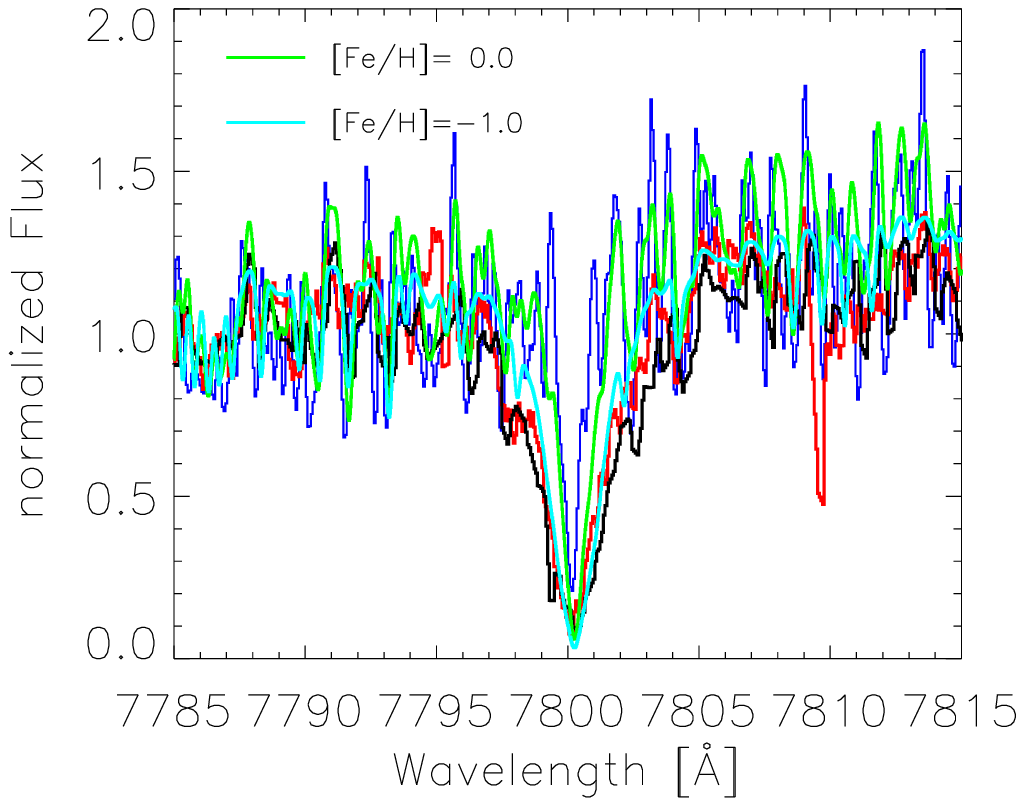}
    \includegraphics[width=.4\textwidth,bbllx=0,bblly=65,bburx=300,bbury=310]{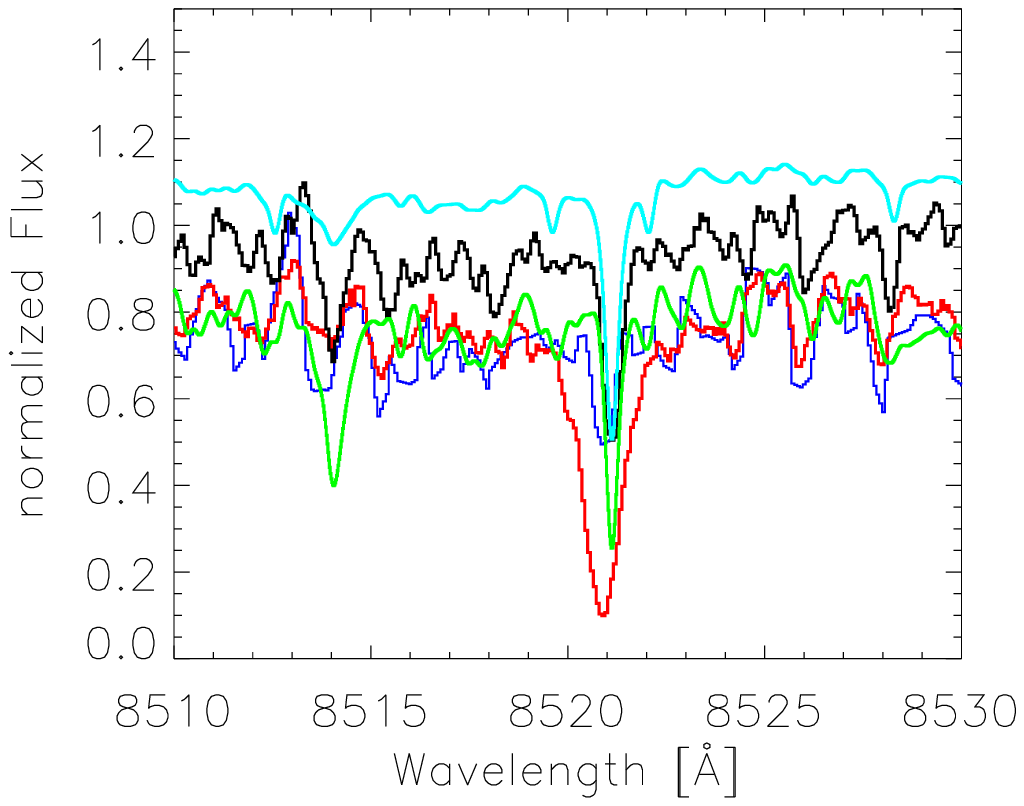}
  }
  \caption{\label{fig:LSR1610_3}Rb and Cs lines as shown in Fig.\,\ref{fig:LSR1610_1}. Two models at $T_\mathrm{eff} = 2800$\,K are overplotted; solar metallicity [Fe/H] = 0.0 (green) and metal deficient [Fe/H]=-1.0 (cyan).}
\end{figure*}

\lsr\ was proposed as an early-type L-subdwarf by L03. They find its
spectral energy distribution to be completely inconsistent with a
spectral type sdM6.0, which would be the spectral classification on
the basis of the standard classification scheme developed by
\cite{Gizis97} and expanded by \cite{Lepine03a}.  On the other hand,
CV05 find a good overall fit with a normal M6 dwarf.  A second strong
argument for a spectral type early L are the Rb~I lines, which are
atypically strong for a normal M dwarf. We show several spectral
orders of our high resolution spectrum of \lsr\ in
Figs.\,\ref{fig:LSR1610_1} and \ref{fig:LSR1610_2}. As a comparison,
we overplot the spectra of the M-dwarf Gl\,406 (M5.5) and the
early-type L-dwarf 2MASS~1439+1929 (L1.0) that we obtained during the
same night. The most interesting spectral features in ultra low-mass
objects and metal-poor subdwarfs are the metal-oxides, the
metal-hydrides, and the strongly pressure-broadened atomic lines of
alkali and alkali earth elements.  In the following, we discuss the
relevant features that are visible in our spectra.

\subsubsection{Metal oxides}

The metal oxides TiO and VO are the main opacity sources in late
M-stars.  Our spectrum covers three bandheads of TiO, the
$\gamma$-system at 7050\,\AA, the $\epsilon$-system at 8430\,\AA, and
the infrared $\delta$-system at 8870\,\AA. TiO bands are observed to
become stronger with cooler effective temperature among early to
mid-type M-stars. In the field dwarfs, the TiO bands in late M-dwarfs
show a reversal, the $\gamma$-band reaches a maximum near M6.5, the
$\epsilon$-band has its maximum a little later at M7.5
\citep{Mohanty04}.  The weakening of TiO band-strengths in very late
M-dwarfs and early L-dwarfs can be explained by the formation of more
metal-rich condensate minerals \citep[eg.]{Lodders02}, i.e. by the
formation of dust in metal-rich field dwarfs. Furthermore, dust
opacity increases with shorter wavelength and the effects of dust
should be visible at bluer wavelengths first.

The reddest TiO band at 8900\,\AA\ is displayed in the first panel of
Fig.\,\ref{fig:LSR1610_2}. The absorption in \lsr\ is consistent with
the one in Gl\,406 and maybe a little weaker than in \twomfl, although
this band does not show a clear structure and differences between the
three spectra are small. The $\epsilon$-bands at 8450\,\AA\ appear
very similar in the field M- and L-dwarfs since they lie on both sides
of the reversal. The $\epsilon$-band of \lsr\ is much weaker,
indicating metal deficiency or a temperature outside the M5.5--L1
region. The third TiO band at 7050\,\AA, the $\gamma$-band shown in
the upper right panel of Fig.\,\ref{fig:LSR1610_1}, is very strong in
\lsr. It is consistent with the strong absorption observed in Gl\,406
and much weaker than in \twomfl.  The weakness of this band in the
L1-dwarf is likely due to the onset of dust formation. The band is
severely saturated in the two other spectra. At such strong
saturation, the underlying opacity could be different in \lsr\ and
Gl\,406 while producing the same appearance in the $\gamma$-band.
Thus, the TiO bands are not inconsistent with the hypothesis, that
\lsr\ is a metal-deficient M-dwarf.

Our spectrum also includes the VO band at 7900\,\AA, and we confirm
the finding by L03 that VO is not strong in \lsr. In Gl\,406 this 
band is comparably weak, and even in \twomfl\ VO is not particularly 
strong. This means that no clear conclusions about the temperature 
or metallicity of \lsr\ can be drawn from the VO bands. 
CV05 discuss a possible new set of bands of TiO (and
perhaps H$_2$O) at 9200-9400\,\AA\ (their Fig. 6), but suggest that
higher resolution spectra are needed for confirmation. We see no
evidence for TiO in this region, only a lot of telluric lines.

\subsubsection{Metal hydrides}
\label{sect:lsrhydrides}

In metal poor M-dwarfs, metal hydrides are observed to be much
stronger than metal oxides, which is likely due to the competition
with H$_2$O for oxygen \citep{Mould76}. In \lsr, the hydride bands are
comparable to those in Gl\,406. For example, the CaH band at 6800\,\AA\ is
comparable to the one in Gl\,406 (upper left panel in
Fig.\,\ref{fig:LSR1610_1}). The FeH band at 9900\,\AA\ is also of 
comparable strength in the last panel of Fig.\,\ref{fig:LSR1610_2}. These
results imply that \lsr\ is not very metal-poor compared to Gl\,406.

In \twomfl\ (L1), one expects the FeH band at 9900\,\AA to be stronger because
the star is cooler, and this is seen in last panel of Fig.\,\ref{fig:LSR1610_2}.
On the other hand, the upper left panel of Fig.\,\ref{fig:LSR1610_1} shows 
that the individual CaH features are weaker in \twomfl\ than in the
other two objects as is typically observed (although the reason for this is 
not really known). We note that the normalization in this panel is local; 
the right upper panel shows that if the spectrum is normalized at 7050\,\AA, 
then the L1 spectrum lies above the other two. This likely indicates that 
the background opacity (TiO?, VO?) is weaker in the L dwarf than in \lsr, 
probably due to dust formation in the former. 

\cite{Andersson03} suggest that absorption features of TiH could
become visible in spectra of ultralow mass stars, and
\cite{Burgasser04b} claim a first detection of TiH around 9400\,\AA\ 
in \twom\ (see Sect.\,\ref{sect:spectrum_twom}). We compare the
spectra of \lsr\ and \twomfl\ in the second panel of
Fig.\,\ref{fig:LSR1610_2}.  In this spectral order, we did not adjust
the two spectra for their radial velocities in order to match the
telluric H$_2$O features that are visible in that wavelength region.
We included a telluric reference spectrum from HR~153 (spectral type
B2) demonstrating that in \lsr\ all significant absorption features in
that wavelength region are due to telluric features. Individual TiH
lines from the $^4\Phi - X^4\Phi$ band are indicated
\citep{Andersson03}. No extra absorption due to TiH is visible in the
spectrum of \lsr\ (9380--9407\,\AA).

\subsubsection{Atomic lines}
\label{sect:lsratomic}

The only strong atomic lines commonly found in spectra of ultralow
mass stars are lines of neutral alkali elements, as Na~I, K~I, Rb~I
and Cs~I, and some alkali earth elements. Such atoms are not captured
in refractory grains and remain in the cool atmospheres of L-dwarfs.
By far the strongest lines are the ones due to the Na~I and K~I
resonance doublets. Our observations unfortunately do not cover the
Na~I doublet at 8190\,\AA, but CV05 show they are strong.  Our spectra
do cover one of the K~I resonance lines at 7700\,\AA, and this K~I
line in \lsr\ is shown in comparison to Gl\,406 (M5.5) and \twomfl\ 
(L1.0) in the second panel of Fig.\,\ref{fig:LSR1610_1}.  The K~I line
of \lsr\ falls between the two comparison spectra; it is significantly
stronger than in Gl\,406 and it is also stronger than in any field
M-dwarf spectrum we have observed.

The suggestion of an early L-type spectral class of \lsr\ comes in part
from the Rb~I lines. According to L03, their strengths are the
main difference between the spectrum of \lsr\ and that of the sdM8.0
star LSR~1425\,+\,7102. Two strong lines of Rb~I are covered by our
observations, and they are plotted in the second and third panels of
Fig.\,\ref{fig:LSR1610_1}.  Both lines are observed at high SNR and
accurately resemble the shape of the Rb~I lines in \twomfl, while they
are much stronger than in Gl\,406. During our observing run we also
observed two dwarfs of spectral type M9, LHS~2065 and LHS~2924. Both
show Rb~I lines significantly weaker than those observed in \lsr\ and
\twomfl.  Thus, the behavior of the atomic lines of K~I and Rb~I in
our high-resolution spectrum confirms that these lines appear typical 
of a field early-L dwarf.

The observed line strengths of the alkali elements K~I and Rb~I (and
the strong lines of Na~I visible in low resolution spectra of CV05)
lead us to expect strong lines of Cs~I as well. Several Cs~I
lines are covered by our spectrum; the two lines at 8520\,\AA\ and
8944\,\AA\ are shown in the last panel of Fig.\,\ref{fig:LSR1610_1}
and the first panel of Fig.\,\ref{fig:LSR1610_2}, respectively. Both
lines are remarkably weaker than in \twomfl; the one at 8520\,\AA\ 
is about half as strong as in \twomfl, while the one at 8944\,\AA\ 
is hardly detected at all. This behavior is not consistent with a 
an early-L object. 

On the other hand, the Cs~I line is stronger than in Gl\,406, though 
not as comparatively strong as the Rb~I lines. Both Cs~I lines are 
located directly inside TiO-bands, which are weakened at lower 
metallicity. The reduced TiO opacity should enhance the visibility 
of atomic lines. That explanation is supported by the presence of 
a relatively strong line from Ba~I at 8560\,\AA\ (another line of 
Ba~I is detected at 7912\,\AA\ inside a VO band). Most interesting is 
the clear detection of two of the Ca~II triplet lines (at 8498 and 
8542\,\AA). All these lines with strengths greater than in Gl\,406 
(which doesn't really show them at all). The triplet lines require
photospheric conditions hot enough to produce ionized calcium, and 
tend to disappear at M6 except in very active stars (Gl\,406 itself 
is reasonably active, and yet does not show them). Taken together, 
these results are consistent with a mid-M effective temperature, 
but with weakened background opacity due to a lower metallicity.

Our spectrum also covers the wavelength region around 9600\,\AA, where
B03 and \cite{Burgasser04a} report the detection of a so far
unidentified absorption feature. We show that region in the third
panel of Fig.\,\ref{fig:LSR1610_2}.  In the spectrum of \lsr, six
individual absorption features appear between 9580\,\AA\ and
9710\,\AA.  The shape of the features clearly suggests that the
absorption is due to atomic lines, and we identify all six to be due
to absorption of Ti~I (a5F-z5Fo).  The reason why these lines of Ti~I
can be seen probably is their particularly low ground state energy.
The energy levels of the identified lines are taken from \emph{The
  Atomic Line List v2.04}\footnote{\ttfamily
  http://www.pa.uky.edu/$\sim$peter/atomic/}; they are given in
Table\,\ref{tab:Ti}.

\begin{deluxetable}{ccc}
  \tablecaption{\label{tab:Ti}Infrared Ti~I lines, transition type E1,
    term designation is a5F-z5Fo for all lines} \tablewidth{0pt}
  \tablehead{Wavelength in Air & J--J & level energies [cm$^{-1}$]}
  \startdata
  9599.55\,\AA   &   3-4   &  6661.00 -- 17075.31\\
  9638.26\,\AA   &   5-5   &  6842.96 -- 17215.44\\
  9647.37\,\AA   &   2-3   &  6598.75 -- 16961.42\\
  9675.50\,\AA   &   4-4   &  6742.76 -- 17075.31\\
  9688.80\,\AA   &   1-2   &  6556.83 -- 16875.19\\
  9705.68\,\AA   &   3-3   &  6661.00 -- 16961.42\\
  \enddata
\end{deluxetable}

\subsection{\twom}
\label{sect:spectrum_twom}

\begin{figure*}
  \centering
  \leavevmode
  \includegraphics[width=.4\textwidth,bbllx=0,bblly=65,bburx=300,bbury=310]{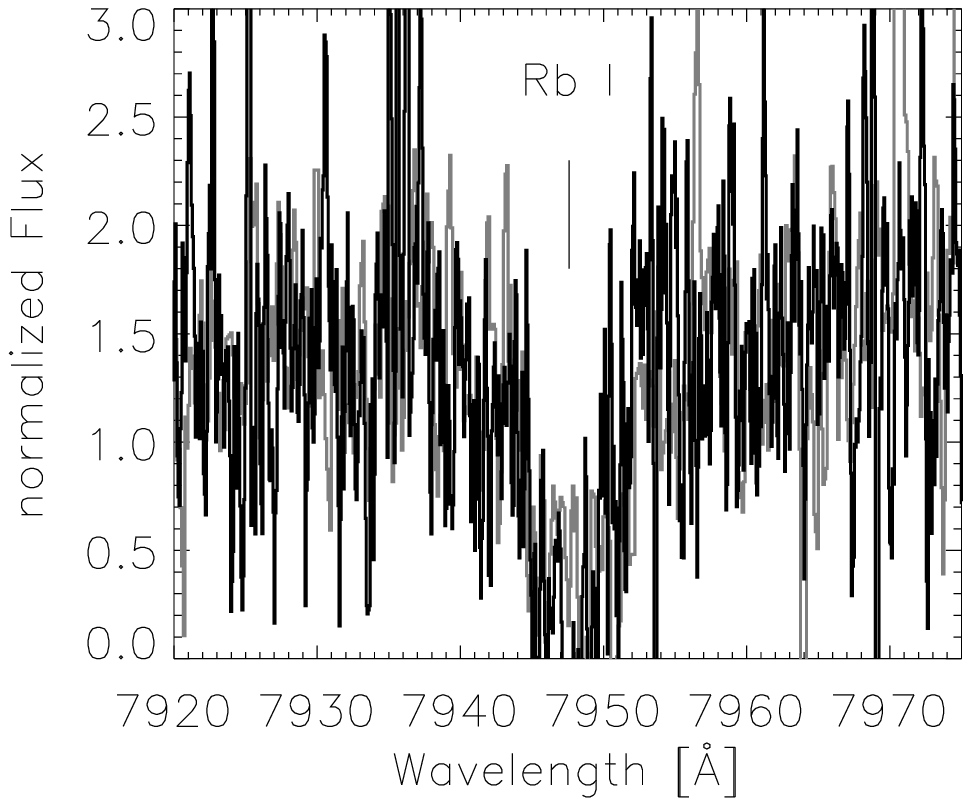}
  \includegraphics[width=.9\textwidth,bbllx=0,bblly=65,bburx=648,bbury=310]{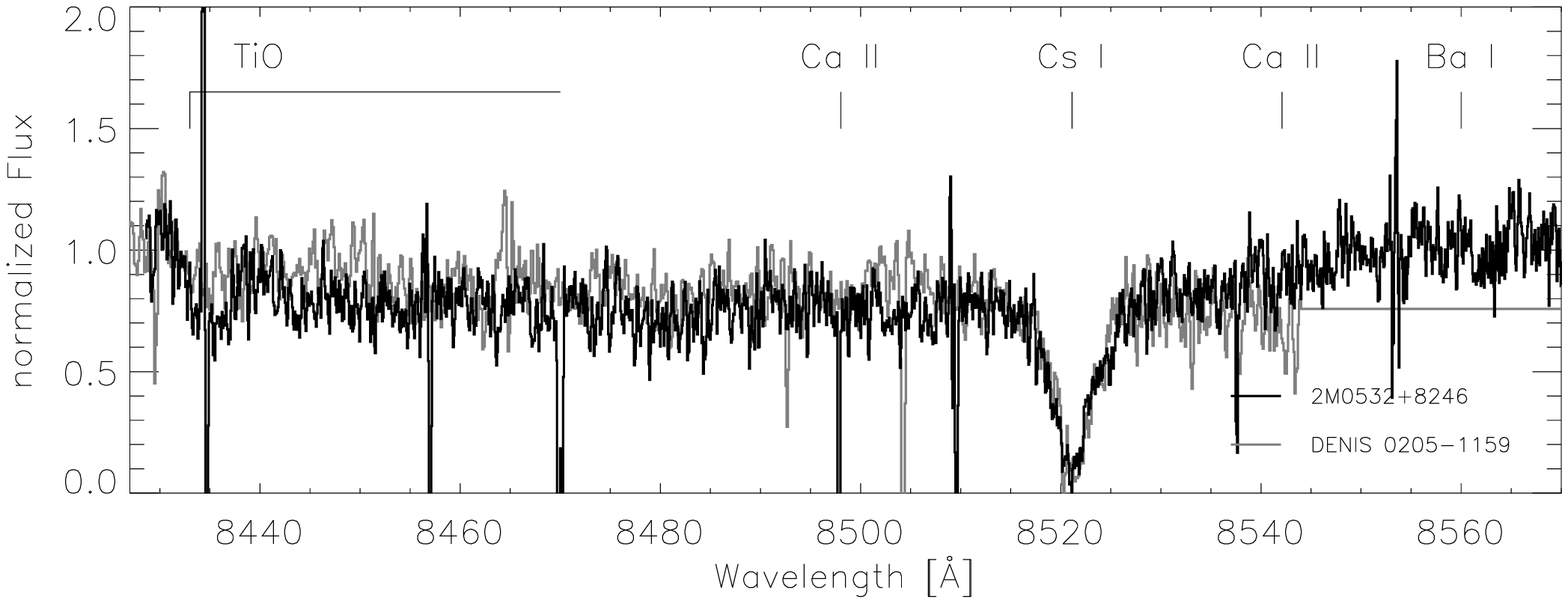}
  \caption{\label{fig:2MASS0532_rb}Spectra of the late-type L-subdwarf \twom\ (black line) compared to the L7 field dwarf \denis\ (grey line).}
\end{figure*}

\begin{figure*}
  \centering \leavevmode
 
\includegraphics[width=.9\textwidth,bbllx=0,bblly=65,bburx=648,bbury=310]{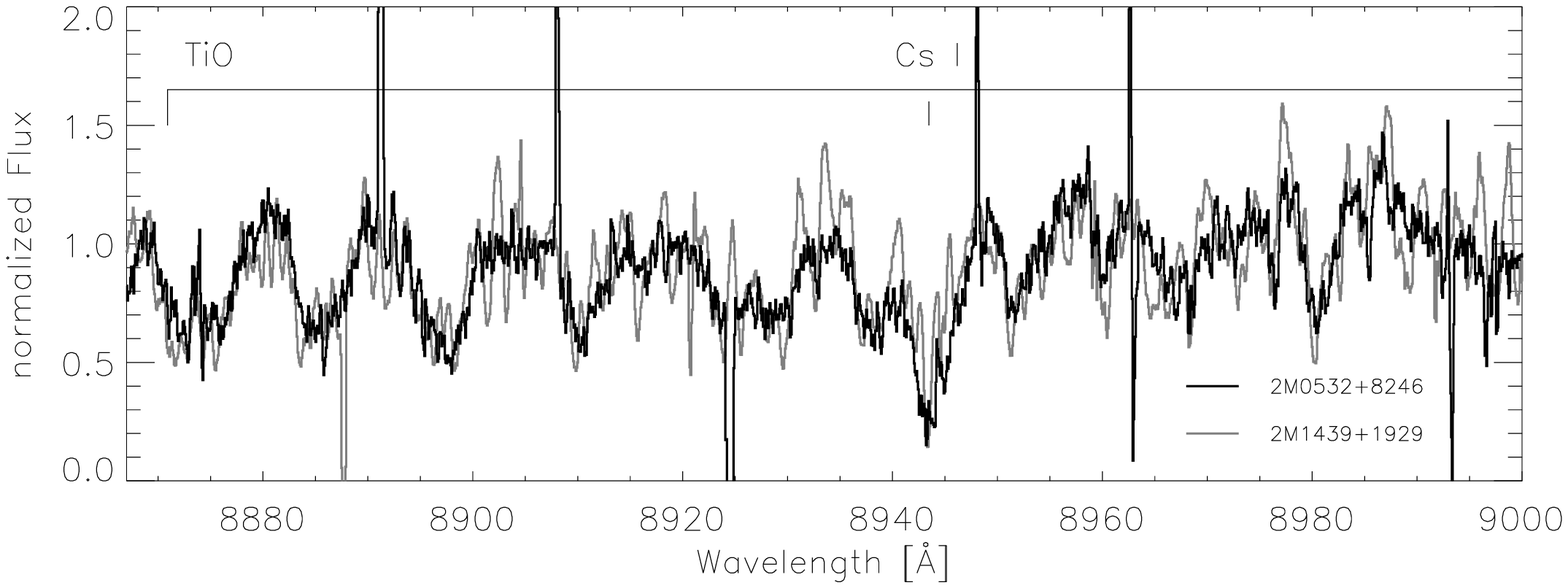}

\includegraphics[width=.9\textwidth,bbllx=0,bblly=65,bburx=648,bbury=310]{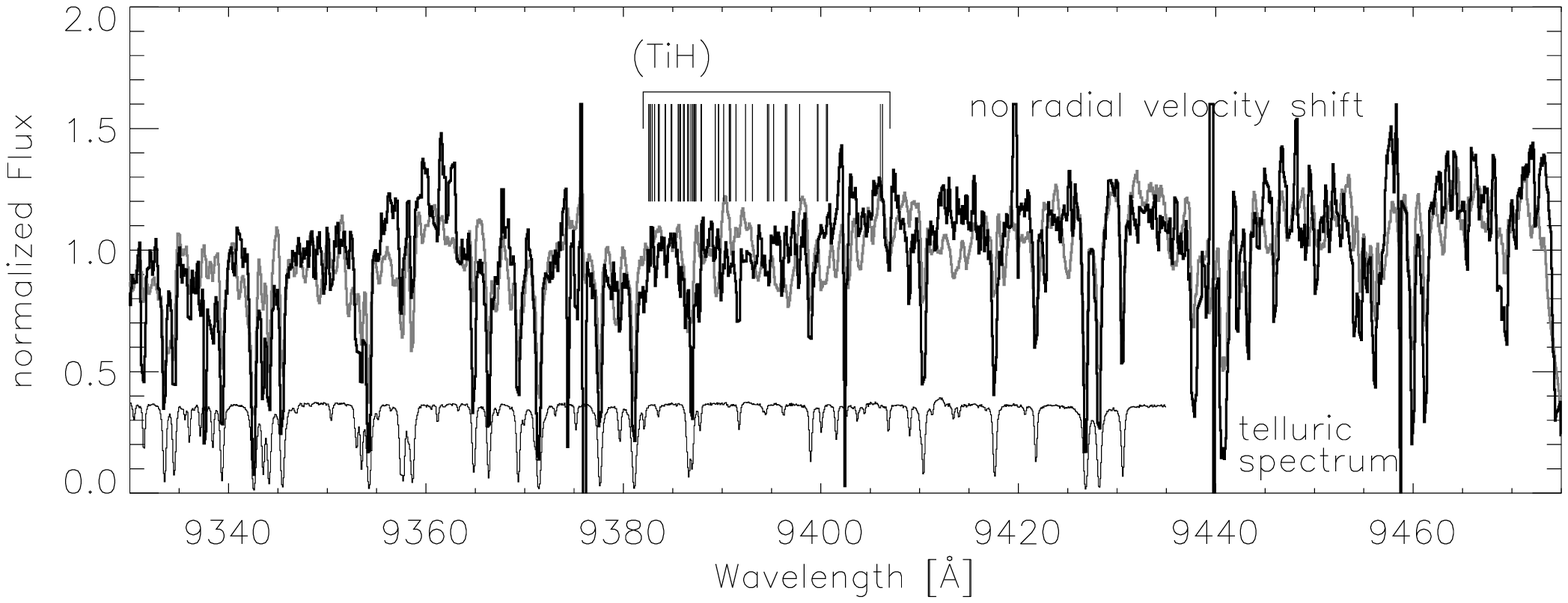}

\includegraphics[width=.9\textwidth,bbllx=0,bblly=65,bburx=648,bbury=310]{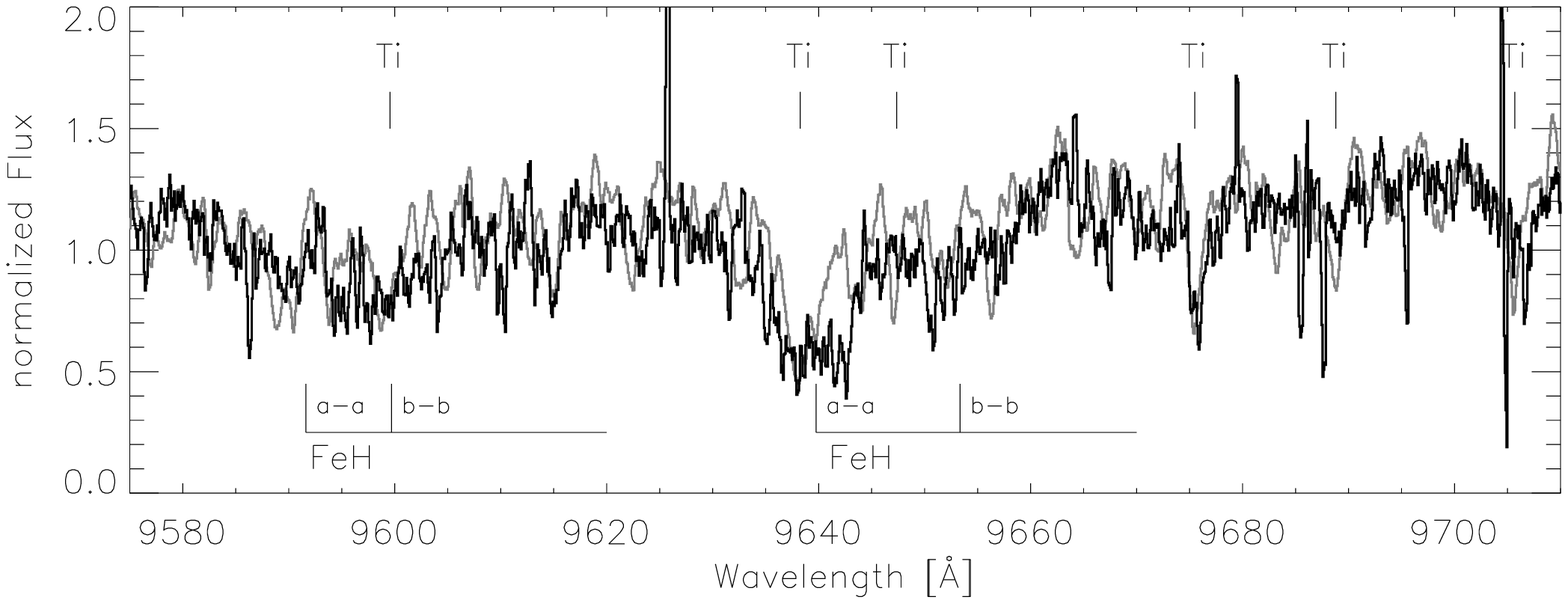}

\includegraphics[width=.9\textwidth,bbllx=0,bblly=65,bburx=648,bbury=310]{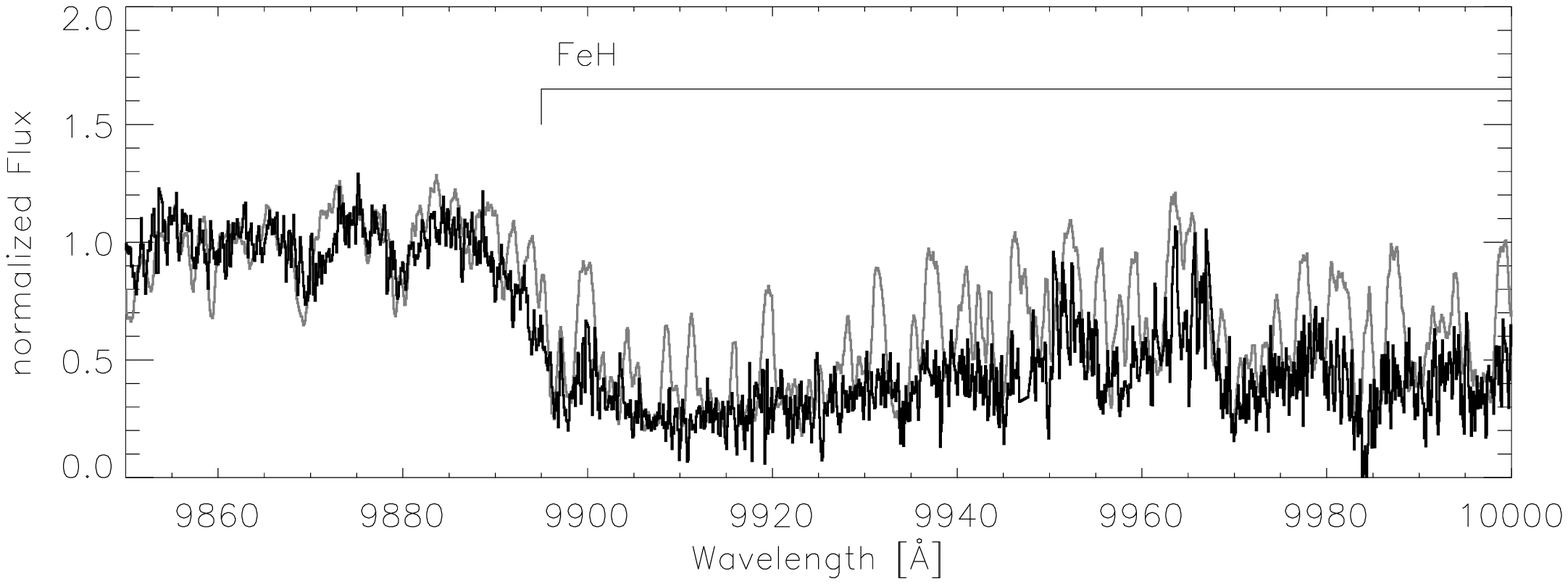}

  \caption{\label{fig:2MASS0532_1}Spectra of the late-type L-subdwarf \twom\ (black line) compared to the L1 field dwarf \twomfl\ (grey line). Our spectrum of \denis\ does not extend to this region.}
\end{figure*}

The optical and infrared spectrum of \twom\ has been investigated by
B03. The $1.5 - 2.4\,\mu$m region is quite unlike any known late-type
M, L, or T-dwarf spectrum, which B03 attribute to collision induced
H$_2$ absorption centered near 2.5\,$\mu$m, enhanced due to
metal-deficiency. The optical and $J$-band region they show, however,
is very similar to that of the L7 dwarf \denis, and B03 suggest that
\twom\ is a metal-poor late L-dwarf similar to \denis. The main
differences with \denis\ are the bands of metal hydrides and TiO,
which are stronger in \twom\ consistent with a metal-poor atmosphere.
B03 suggest that the TiO bands are enhanced because production of
Ti-bearing condensates is inhibited in the Ti depleted atmosphere,
leaving Ti and O for TiO.
 
The high-resolution spectrum we obtained of \twom\ is of lower quality
than the one shown in the previous Section for \lsr. Nevertheless, in
the red part of the spectrum, data quality is high enough to compare
the Rb~I and Cs~I lines redward of 8700\,\AA\ to a spectrum of \denis\ 
we took with HIRES at Keck in December~1997. We plot parts of both
spectra in the two panels of Fig.\,\ref{fig:2MASS0532_rb}. The
spectrum we have of \denis\ only extends to 8700\,\AA. The other parts
of our high resolution spectrum of \twom\ are plotted together with
the spectrum of the L1 field dwarf \twomfl\ in
Fig.\,\ref{fig:2MASS0532_1}.

\subsubsection{Metal oxides}

In M-type subdwarfs, absorption bands of metal oxides like TiO and VO
are weaker than they are in M-dwarfs of higher metallicity. TiO bands
become saturated in mid M-dwarfs then weaken in late M- and L-dwarfs
due to condensation of Ti into refractory dust grains.  The TiO
band at 8430\,\AA\ is stronger in \twom\ than it is in \denis, as
mentioned by B03. This implies that the formation of dust is more 
inhibited than the formation of TiO in the metal-poor atmosphere of \twom.
The TiO band at 8870\,\AA\ is compared to the L1 dwarf \twomfl\ in the
top panel of Fig.\,\ref{fig:2MASS0532_1}. Both have comparably strong
TiO bands, i.e., the TiO band of \twom\ is as strong as is in a
metal rich L1 dwarf, and much stronger than in a normal metallicity 
dwarf of spectral type L7.

\subsubsection{Metal hydrides}

The FeH and CrH bands at 9900\AA\ and 9970\AA, respectively, are
stronger than in any other late type dwarf known to date (bottom panel 
of Fig.\,\ref{fig:2MASS0532_1}). Their strength can be interpreted as a 
direct consequence of the enhanced hydride band strength in 
metal-poor atmospheres.

A detection of TiH at 9400\,\AA\ in the spectrum of \twom\ shown in
B03 was claimed by \cite{Burgasser04b}, and \cite{Burrows05} showed
that spectral features of TiH at this wavelengths may not be weak.
Strong TiO absorption bands together with strongly enhanced bands of
FeH and CrH support the expectation that \twom\ is a very good
candidate for a detection of TiH absorption. We show the region around
9400\,\AA\ of our spectrum of \twom\ compared to the spectrum of
\twomfl\ and the telluric reference spectrum of HR~153 in the second
panel of Fig.\,\ref{fig:2MASS0532_1}.  We did not adjust the spectra
for their radial velocities, in order to show the matching sets of
telluric lines due to H$_2$O. There is no positive evidence of TiH in
the 9380--9407\,\AA\ region of our high resolution spectrum. All
visible absorption features are narrow and clearly not rotationally
broadened.  They also match the features visible in \twomfl\ and the
telluric reference. Thus, TiH absorption is not detected in the
high-resolution spectrum of \twom.

\subsubsection{Atomic lines}

The spectral quality in the blue part of the spectrum of \twom\ is not
very high. We cannot reliable follow the spectrum around the K~I line
at 7700\,\AA. The bluest detected spectral feature is the Rb~I line at
7950\,\AA. Another atomic alkali line clearly detected is the Cs~I
line at 8520\,\AA. Both regions are shown together with the spectrum
of the L7 dwarf \denis\ in Fig.\,\ref{fig:2MASS0532_rb}. The strong
alkali lines of \denis\ are accurately reproduced by the spectrum of
\twom, possibly indicating comparable temperatures in both objects.

No lines of Ca~II or Ba~I are detected in \twom. Since these lines are
expected to be comparably weak, they can be easily hidden in the noise
since they would be broadened by the strong surface rotation.  Cs~I is
also detected at 8940\,\AA\, but we have no comparison spectrum of
\denis\ in that region. Comparison to \twomfl\ shows that Cs~I is much
stronger than it is in the L1 dwarf, as expected.

We show the spectral region around 9640\,\AA\ in the third panel of
Fig.\,\ref{fig:2MASS0532_1}. B03 detected an unidentified absorption
feature in that region. We have shown in Sect.\,\ref{sect:lsratomic}
that atomic Ti absorption lines are remarkably strong in that spectral
region in \lsr. The major absorption features visible in the spectrum
of \twom\ are located at the same wavelengths, and we conclude that a
great part of the unidentified absorption is due to atomic Ti.

The strongest absorption line at 9640\,\AA\ appears to have a
very strong red wing; this cannot be attributed to the fast rotation
velocity which only marginally broadens the line compared to the
intrinsic width visible in the spectra of \lsr\ and \twomfl. We looked
at synthetic spectra at this spectral region and found that this red
wing also appears in spectra at temperatures around 2000\,K. This
absorption is probably due to lines of FeH which cover the region from
9590 to 9700\,\AA. They particularly show a cluster of lines at
9643\,\AA\footnote{We thank P.H. Hauschildt for investigating
  absorption features in PHOENIX model spectra.}, which we identified
as a (0--0) band, particularly the ${^4\Delta_{3/2}}-{^4\Delta_{5/2}}$
transition \citep{Phillips87}.  Since FeH is particularly strong in
the spectrum of \twom, it appears plausible that FeH shows up in this
spectral region as well, although it has not yet been observed in
other L-dwarfs. A second FeH bandhead, the
${^4\Delta_{5/2}}-{^4\Delta_{7/2}}$ transition, may also be visible at
9600\,\AA. Both bandheads split into two different parities (a--a) and
(b--b). We marked them in the third panel of
Fig.\,\ref{fig:2MASS0532_1}.

\section{Resolving the Alkali Line Conundrum in \lsr}
\label{sect:lsrmodels}

We return now to the apparent inconsistencies in the spectrum of \lsr,
and the question of whether it is really M or L, and metal-poor or not.
We first compare the appearance of the Rb~I and Cs~I lines in \lsr\ 
to spectra of Gl\,406 (M5.5V) and \twomfl\ (L1). We wish to investigate 
why the Rb~I lines of \lsr\ resemble the lines \twomfl\ while the 
Cs~I lines of \lsr\ resemble those of Gl\,406.  
We compare the spectra of \lsr\ and \twomfl\ 
to synthetic spectra calculated with the PHOENIX code \citep[]
[we use the DUSTY models and assume log\,$g = 5.5$ for both objects]
{Allard01}. The temperature scale in \cite{Golimowski04} produces an 
estimate of the temperature of $T_{\rm eff} \approx 2200$\,K for 
\twomfl\ (L1), and we use a temperature of 2800\,K for comparison to
Gl\,406. We adopt the hypothesis that this is the temperature appropriate
to \lsr, and test models with metallicities of [Fe/H]~=~0 and $-1$.

In the left panel of Fig.\,\ref{fig:LSR1610_3} we show the three 
observed spectra and two models for the Rb~I line at 7800\AA. The 
lines in \lsr\ and \twomfl\ are very similar (\lsr\ has a slight broader
one). Gl\,406 has the weakest, narrowest line. The model with solar
metallicity most closely resembles Gl\,406; although the line depth is
deeper the line wings end at almost the same width. The metal-deficient
model, on the other hand is as deep as \lsr\ and \twomfl, and has 
wings which are broad, but not quite broad enough to match them.
The molecular pseudo-continua around the line also match the two cases.
We caution that the models still have deficiencies, and the normalization
of the spectra is a matter of (consistent) choice, but the general 
agreements are heartening. 

The right panel of Fig.\,\ref{fig:LSR1610_3} shows a similar analysis
for the Cs~I line at 8521\AA. The spectra are all normalized as in Fig.
\,\ref{fig:LSR1610_1}, namely outside the TiO band at 8430\AA. The 
great difference between Gl\,406 and \twomfl\ in Cs~I is quite obvious,
although the similar strengths of the background TiO leaves their pseudo-
continua at similar levels. The solar metallicity model has a similar
level, and the line is sharper and deeper than Gl\,406, but nothing like
the strength of the line for the L dwarf. Lowering the metallicity yields
a line strength and pseudo-continuum level that is a better match to \lsr.
We thus obtain a reasonable fit to both the Rb~I \emph{and} Cs~I lines 
of \lsr\ using approximately $T_{\rm eff} = 2800$\,K, [Fe/H]~=~$-1$. 
This is why we would assign it a spectral type of sd?M6. The reason for 
the disparate line strengths in Gl\,406 compared with \lsr\ is 
the weakened background opacity in \lsr, which allows the line wings of
Rb~I (but not Cs~I, which is less abundant) to be revealed, while they
remain hidden by TiO in Gl\,406. This is also the reason for the increased
strengths of the Ba~I and Ca~II lines in \lsr\ compared to Gl\,406.

\section{Conclusions}

We have presented the first high resolution spectra of the peculiar
possibly metal-poor sd?M6 \lsr\ and the late-type L-subdwarf
\twom. We have derived accurate radial and rotation velocities for
both objects, and we compared their spectral features to spectra of
known field L-dwarfs. The spectra of \lsr\ and \twom\ are very 
different in many respects. These are discussed below.

The radial velocity of \lsr\ is consistent with the value reported by
L03 if we assume that their determination suffered similar systematic
effects as mentioned by B03 for the case of \twom. For the latter, we
derive a radial velocity consistent with the result of B03. We
computed the space motion in order to check membership of the halo
population. Our result for \lsr\ significantly differs from the result
reported by L03; the radial velocity of \lsr\ does not fall outside
the $2\sigma$-region reported for thick disk stars by \cite{Chiba00}.
However, the $V$-velocity of \lsr\ ($-111$\,km\,s$^{-1}$) is
comparably high, and the space motion vector does not contradict its
halo membership. For \twom, our estimate of the $V$-velocity is $V =
-285$\,km\,s$^{-1}$, which is in strong support of halo membership.

The rotation velocity of \lsr\ is below the detection limit of about
5\,km\,s$^{-1}$. This is consistent with the bulk of its properties
more closely resembling the M5.5V dwarf Gl\,406 than an L1 object.
We have analyzed the reasons why it was initially classified as an
early-L subdwarf, and explain them with a mild metal-deficiency.
For the true L subdwarf, \twom, we measure a rotation velocity of
$v\,\sin{i} = 65$\,km\,s$^{-1}$. If this object is indeed a relic of
the young galaxy, that means that the braking time for a mid L-dwarf
is much longer than 10\,Gyr. Either the braking time exceeds the
age of the galaxy by this spectral type, or there is no rotational
braking in mid to late L-type objects at all. 

In M-subdwarfs metal-oxides are weaker and metal-hydrides are
stronger than in the metal rich M-dwarfs, but both species are 
stronger in \twom\ than in its metal-rich counterpart \denis. 
The strong oxide-bands in \twom\ can be explained by invoking 
a reduced level of dust formation in the metal-deficient atmosphere, 
which would leave behind more TiO than in the metal rich L-dwarfs (B03). 
The extremely strong metal-hydride bands in \twom\ are generally
consistent with observations of enhanced metal-hydride bands in
M-subdwarfs.

Chemistry and line formation in the ultracool atmospheres of L-dwarfs
is much more complex than it is in hotter stars because of dust
formation and depletion of refractory metals from the atmosphere. The
chemical anomalies of subdwarfs heighten these complexities, and
currently yield a confusing situation.  High-resolution spectra
provide crucial information for our eventual understanding of the 
ultracool metal-deficient atmospheres which were common in the 
early galaxy.

\acknowledgments Based on observations obtained from the W.M. Keck
Observatory,which is operated as a scientific partnership among the
California Institute of Technology, the University of California and
the National Aeronautics and Space Administration. We would like to
acknowledge the great cultural significance of Mauna Kea for native
Hawaiians and express our gratitude for permission to observe from
atop this mountain. We thank P.H. Hauschildt for providing
high-resolution versions of spectra calculated with PHOENIX, and for
an identification run on some of them. GB thanks the NSF for grant
support through AST00-98468. AR has received research funding from the
European Commission's Sixth Framework Programme as an Outgoing
International Fellow (MOIF-CT-2004-002544).

%% copy editing.  Individual instruments can be provided in parentheses,
%% after the keyword, but they will not be verified.
%Facilities: \facility{Nickel}, \facility{HST(STIS)}, \facility{CXO(ASIS)}.
%\appendix

 \end{document}